\documentclass[conference]{IEEEtran}
\IEEEoverridecommandlockouts
\usepackage{cite}
\usepackage{amsmath,amssymb,amsfonts}
\usepackage[ruled,vlined]{algorithm2e}
\usepackage{graphicx}
\usepackage{textcomp}
\usepackage[table]{xcolor}
\usepackage{hyperref}
\usepackage[nolist]{acronym}
\usepackage{tabularx}
\usepackage{booktabs}
\usepackage{soul}
\usepackage{fancyhdr}
\fancypagestyle{firststyle}{
    \fancyhf{}
    \fancyhead[L]{D. Wang, I. Jariwala, A. Bazzi, S. Rangan, T. S. Rappaport, and M. Chafii,``Compressed Multiband Sensing in FR3 Using Alternating Direction Method of Multipliers,'' accepted by \textit{IEEE Wireless Communications and Networking Conference (WCNC)}, Kuala Lumpur, Malaysia, Apr. 2026, pp. 1--6.}
}

\begin{acronym}
    \acro{AWGN}{additive white Gaussian noise}
    \acro{ISAC}{integrated sensing and communication}
    \acro{DFRC}{dual-function radar communication}
    \acro{AoA}{angle of arrival}
    \acrodefplural{AoA}[AoAs]{angles of arrival}
    \acro{ToA}{time of arrival}
    \acrodefplural{ToA}[ToAs]{times of arrival}
    \acro{AoD}{angle of departure}
    \acrodefplural{AoD}[AoDs]{angles of departure}
	\acro{LoS}{line-of-sight}
	\acro{ULA}{uniform linear array}
	\acro{OFDM}{orthogonal frequency-division multiplexing}
	\acro{CRB}{Cram\'er-Rao bound}
	\acro{i.i.d}{independently and identically distributed}
	\acro{DFT}{discrete Fourier transform}
	\acro{SISO}{single-input, single-output}
    \acro{SIMO}{single-input, multiple-output}
    \acro{MIMO}{multiple-input, multiple-output}
	\acro{MSE}{mean squared error}
	\acro{SNR}{signal-to-noise ratio}
	\acro{DMC}{dense multipath component}
	\acro{SC}{specular component}
	\acro{RCS}{radar cross section}
	\acro{TF}{transfer function}
	\acro{PDP}{power-delay profile}
	\acro{PAP}{power-angular profile}
	\acro{WB}{wideband}
	\acro{UWB}{ultra-wideband}
	\acro{SV}{Saleh-Valenzuela}
	\acro{AS}{angular spread}
	\acro{DS}{delay spread}
	\acro{FCF}{frequency correlation function}
	\acro{VMD}{Von-Mises distribution}
	\acro{FIM}{Fisher information matrix}
    \acro{ADMM}{alternating direction method of multipliers}
    \acro{RMSE}{root mean squared error}
    \acro{MRC}{maximal ratio combining}
    \acro{SRP}{successful recovery probability}
    \acro{CS}{compressed sensing}
    \acro{CMS}{compressed multiband sensing}
    \acro{HPC}{high performance computing}
    \acro{SAGE}{space alternating generalized expectation-maximization}
    \acro{Tx}{transmitter}
    \acro{Rx}{receiver}
    \acro{RE}{resource element}
    \acro{BF}{beamforming}
    \acro{QAM}{quadrature amplitude modulation}
\end{acronym}

\definecolor{gray10pct}{gray}{0.9}

\DeclareMathOperator{\trace}{tr}

\DeclareMathOperator{\ve}{vec}

\DeclareMathOperator{\diag}{diag}
\DeclareMathOperator{\cat}{cat}
\DeclareMathOperator{\prox}{prox}
\DeclareMathOperator{\NF}{NF}

\DeclareMathOperator*{\argmax}{arg\,max}

\newcommand{\norm}[1]{\left\lVert#1\right\rVert}

\def\BibTeX{{\rm B\kern-.05em{\sc i\kern-.025em b}\kern-.08em
    T\kern-.1667em\lower.7ex\hbox{E}\kern-.125emX}}
\begin{document}
\addtolength{\topmargin}{0.05in}

\addtolength{\oddsidemargin}{-0.02in}

\title{Compressed Multiband Sensing in FR3 Using Alternating Direction Method of Multipliers}

\author{
\IEEEauthorblockN{
Dexin Wang\IEEEauthorrefmark{1}\IEEEauthorrefmark{2}, 
Isha Jariwala\IEEEauthorrefmark{2},
Ahmad Bazzi\IEEEauthorrefmark{1}\IEEEauthorrefmark{2}, 
Sundeep Rangan\IEEEauthorrefmark{2},
Theodore S. Rappaport\IEEEauthorrefmark{2},
and Marwa Chafii\IEEEauthorrefmark{1}\IEEEauthorrefmark{2}}

\IEEEauthorblockA{\IEEEauthorrefmark{1}Engineering Division, New York University (NYU) Abu Dhabi, Abu Dhabi, UAE.}
\IEEEauthorblockA{\IEEEauthorrefmark{2}NYU WIRELESS, NYU Tandon School of Engineering, New York, USA}}

\maketitle
\thispagestyle{firststyle} 

\begin{abstract}
Joint detection and localization of users and scatterers in multipath-rich channels on multiple bands is critical for integrated sensing and communication (ISAC) in 6G. Existing multiband sensing methods are limited by classical beamforming or computationally expensive approaches. This paper introduces alternating direction method of multipliers (ADMM)-assisted compressed multiband sensing (CMS), hereafter referred to as ADMM-CMS, which is a novel framework for multiband sensing using uplink quadrature amplitude modulation-modulated pilot symbols. To solve the CMS problem, we develop an adaptive ADMM algorithm that adjusts to noise and ensures automatic stopping if converged. ADMM combines the decomposability of dual ascent with the robustness of augmented Lagrangian methods, making it suitable for large-scale structured optimization. Simulations show that ADMM-CMS achieves higher spatial resolution and improved denoising compared to Bartlett-type beamforming, yielding a 34 dB gain in per-antenna transmit power for achieving a 0.9 successful recovery probability (SRP). Moreover, compared to performing compressed sensing separately on the constituent 7 GHz and 10 GHz sub-bands, ADMM-CMS achieves reductions in delay root mean squared error of 34.46\% and 40.76\%, respectively, at -41 dBm per-antenna transmit power, while also yielding improved SRP. Our findings demonstrate ADMM-CMS as an efficient enabler of ISAC in frequency range 3 (FR3, 7-24 GHz) for 6G systems.
\end{abstract}


\section{Introduction}
\label{sec:introduction}
\Ac{ISAC} has recently been proposed as one of the six usage scenarios in IMT-2030 (6G), and has become a practical necessity for 6G wireless systems \cite{chafii2023twelve}.
To save resources in spectrum, processing power, and hardware, which were traditionally allocated separately for sensing and communications, \ac{ISAC} systems must not only deliver multi-gigabit data rates but also jointly detect and localize objects in the environment with high accuracy \cite{rap2019100}. Thus, \ac{ISAC} systems are envisioned to efficiently foster future applications of 6G networks such as Internet of Things, smart cities, and autonomous driving. In communication-centric \ac{ISAC}, the goal is to perform sensing via existing communication symbols such as pilots. \Ac{CS} provides an elegant solution to \ac{ISAC} problems, offering reduced noise \cite{wu2011csnoisered}, hence suppression of false target detection. 
From a communication-centric \ac{ISAC} perspective, in order to maintain sensing performance, \ac{CS} requires less sacrifice in communication resources such as the \ac{OFDM} \acp{RE} as \ac{CS} enables sub-Nyquist sampling.

The introduction of new frequency bands for cellular mobile communication, such as frequency range 3 (FR3, 7–24 GHz), prompted multiband sensing to become an enabler of ISAC, allowing the ability to take advantage of both the coverage of the lower bands and the spatial resolution of the higher bands  (due to narrrower beams) to improve the sensing accuracy and even help reduce ambiguities, such as grating lobes \cite{Shakya2025icc, shakya2024meas, Ying2025icc,Shakya2024gc2,bazzi2025uppermidbandspectrum6g}. This combination of different bands is enabled by the important property that the angles and delays of the targets remain relatively constant across sub-bands while their amplitudes may vary \cite{shakya2024meas, Shakya2024gc2}. Throughout this paper, we assume sub-band bandwidths in the hundreds of MHz range, consistent with FR3 allocations \cite{bazzi2025uppermidbandspectrum6g}. Within the upper mid-band, angular spread and delay spread remain relatively constant, with only minor frequency dependence. 
Moreover, fragmented spectrum allocations\cite{Ted2025icc, bazzi2025uppermidbandspectrum6g}
in FR3 make multiband sensing more suitable.

Past work on joint detection and localization has relied heavily on classical \ac{BF}. Bartlett and Capon beamformers have been widely studied for localization, but their resolution is limited by array aperture, and they suffer under low-\ac{SNR} conditions \cite{raviv2024multi}. Iterative refinements on the estimated parameters such as the \ac{SAGE} algorithm \cite{Gong2023SAGE, liberti1999smart } 
improve estimation accuracy but at the expense of complexity. 
Approaches not relying on peak-finding on a discrete angle-delay grid, such as atomic norm minimization or subspace methods achieve excellent resolution but remain computationally expensive and difficult to scale in multiband scenarios \cite{chi2015atomicnorm}. Although recent studies in FR3 have begun to explore multiband localization \cite{bazzi2025uppermidbandspectrum6g, raviv2024multi}, most existing approaches continue to rely on BF-based estimators instead of sparse recovery methods such as \ac{CS} \cite{liu2018crb}. In this paper, our contributions are as follows:
\begin{itemize}
    \item We propose \ac{ADMM}-assisted \ac{CMS}, hereafter referred to as ADMM-CMS, which is a new framework of using \ac{CS} to perform multiband \ac{ISAC} on FR3 using quadrature amplitude modulation-modulated communication pilot symbols in the uplink. We formulate the multiband joint angle–delay estimation problem as a structured sparse recovery task.

    \item We develop an \ac{ADMM} algorithm with adaptive constraint and penalty parameters to solve the \ac{CMS} problem efficiently, to adapt to varying noise conditions, and also to enable automatic stopping of the algorithm once it has converged sufficiently.

    \item  We show that in higher \ac{SNR} regimes, our ADMM-CMS approach has lower \ac{RMSE} of localization and better \ac{SRP} compared to \ac{CS} on each constituent sub-band. 
    
    \item We also show that ADMM-CMS offers sharper spatial resolution and less noisy peaks, which translates to higher \ac{SRP}, as compared to Bartlett-type \ac{BF}. The results shown herein indicate the promise of ADMM-CMS as an efficient enabler of \ac{ISAC} in FR3 for 6G systems.

\end{itemize}

\noindent \textbf{Notations}: $\otimes$ denotes the Kronecker product and $\odot$ denotes the element-wise product.
Additionally, $\ve(\cdot)$ denotes reshaping a matrix into a column vector, and $\ve^{-1}(\cdot)$ denotes the inverse operation of $\ve(\cdot)$ .
Also, $\overline{a}$ denotes the complex conjugate of $a$, and $\{ a_n \}_{n=1}^{N}$, which may also be written as $a_{1...N}$, denotes the set $\{ a_1,a_2,\dots,a_N \}$. 
Moreover, $[ \pmb{a}_n ]_{n = 1}^N$ is the vertically stacked vector $[\pmb{a}_1^T,\pmb{a}_2^T,\dots,\pmb{a}_N^T ]^T$, and $\cat_n\{\cdot\}$ denotes concatenating tensors along the $n^{\text{th}}$ dimension. 
Finally, for some matrix $\pmb{A}$, $\left[\pmb{A}\right]_{i,:}$ denotes the $i^{\text{th}}$ row of $\pmb{A}$, $\left[\pmb{A}\right]_{:,j}$ denotes the $j^{\text{th}}$ column of $\pmb{A}$, $\left[\pmb{A}\right]_{i,j}$ denotes the $(i,j)^{\text{th}}$ element of $\pmb{A}$, $\norm{\pmb{A}}_F$ denotes the Frobenius norm of $\pmb{A}$, and $\norm{\pmb{A}}_{a,b}$ denotes the $\ell_{a,b}$-mixed norm of $\pmb{A}$.

\section{System Model}
\label{sec:system-model}

\subsection{Multiband Channel and Signal Model}

\begin{figure}[!t]
\centering
\includegraphics[width=3.5in]{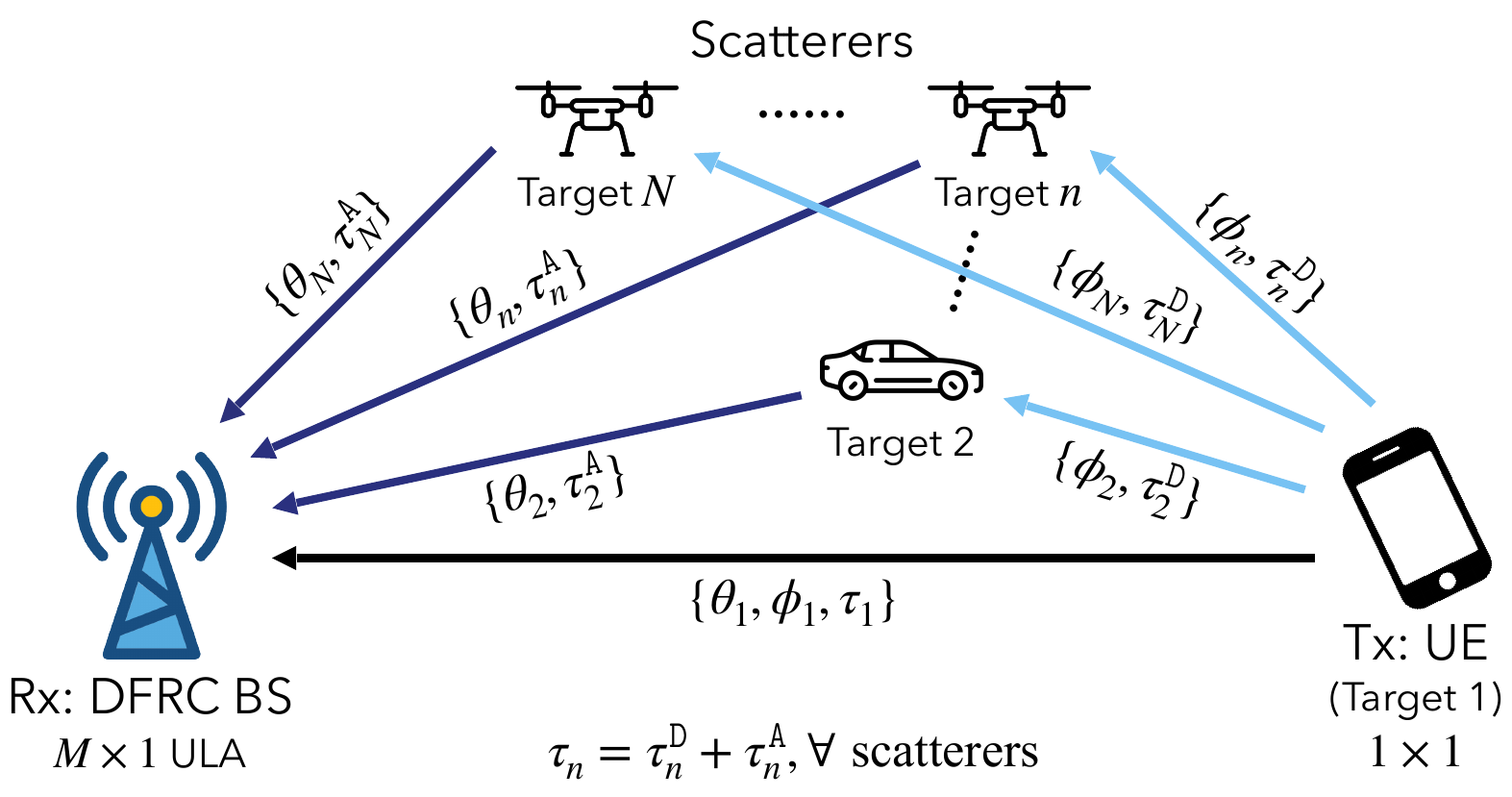}
\vspace{-10pt}
\caption{The cellular \ac{ISAC} scenario with one base station operating in bistatic mode and $N$ targets, including one user device during uplink channel transmission and $N-1$ scatterers within the range of the user device.}
\label{fig:scenario}
\vspace{-10pt}
\end{figure}

As shown in Fig. \ref{fig:scenario}, we consider a reverse channel (e.g. uplink) scenario with one \ac{Rx} operating as a \ac{DFRC} base station in bistatic mode on $K$ sub-bands. There is one transmitting uplink communication user, and the channel consists of $N-1$ scatterers. We consider both the user and the scatterers as \textit{targets}, meaning that we are interested in localizing $N$ targets. The \ac{Rx} has $M$ \ac{ULA} antennas and operates on $Q$ subcarriers per sub-band. We assume that the \ac{Tx}, \ac{Rx}, and the scatterers are stationary, the location and orientation of the \ac{Rx} and the \ac{Tx}-\ac{Rx} clock synchronization error are known, and the \ac{LoS} signal is ideally always received. We can thus localize the targets from estimating the \acp{AoA} and \acp{ToA}.

For the $k^\text{th}$ sub-band, we consider the \ac{SIMO} channel in the frequency domain as follows.
\begin{equation}
    \pmb{H}_k = 
    \sum_{n=1}^N 
    g_{n,k} 
    \bigl(\pmb{a}^{\tt{R}}_k(\theta_n)\bigr)
    \bigl(\pmb{a}^{\tt{F}}_k(\tau_n)\bigr)^T
    \in\mathbb{C}^{M\times Q},
\end{equation}
where $\theta_n$ and $\tau_n$ are the \ac{AoA} and the \ac{ToA} of the $n^\text{th}$ target, respectively, defined from the antenna broadside. The steering vectors are given by
\begin{equation}
    \pmb{a}^{\tt{R}}_k(\theta) = 
    \left[
		\exp\left(
			{-j 2\pi f_k (d_k/c)}m
			\sin\theta
		\right)
	\right]_{m = 0}^{M-1},
\end{equation}
\begin{equation}
    \pmb{a}^{\tt{F}}_k(\tau) = 
    \left[
		\exp\left(
			-j 2\pi (f_k+q\Delta f_k) 
            \tau
		\right)
	\right]_{q = 0}^{Q-1},
\end{equation}
where $m$ is the \ac{Rx} antenna index, $q$ is the subcarrier index, $f_k$ is the sub-band frequency, $\Delta f_k$ is the subcarrier spacing, and $d_k$ is the antenna spacing for the $k^{\text{th}}$ sub-band. Note that our proposed method can be extended to a \ac{MIMO} channel model as well. We assume a simplified scenario where the antenna spacing differs per sub-band with $d_k=\lambda_k/2$, where $\lambda_k=c/f_k$ is the wavelength for sub-band $k$. The gain of the  \ac{LoS} path is modeled by the Friis' equation
\begin{equation}
\label{Friis}
	g_{1,k} = 
    \xi_{1,k}
    \sqrt{
	\frac{P^{\tt{T}} \lambda_k^2u_k^{\tt{R}}(\theta_1)u_k^{\tt{T}}(\phi_1)}
	 {(4\pi)^2 (c\tau_{1})^{\beta_k}}},
\end{equation}
and the gain of the $n^\text{th}$ scatterer is modeled by the two-way bistatic radar equation
\begin{equation}
\label{2waybistaticradar}
	g_{n,k} = 
	\xi_{n,k} 
	\sqrt{\frac{P^{\tt{T}}\lambda_k^2 	u_k^{\tt{R}}(\theta_n) u_k^{\tt{T}}(\phi_n)}
	{(4\pi)^3(c\tau_n^{\tt{A}})^{\beta_k}(c\tau_n^{\tt{D}})^{\beta_k}}},
\end{equation}
where $P^{\tt{T}}$ is the average transmit power, $u_k^{\tt{T}}(\phi)$ and $u_k^{\tt{R}}(\theta)$ are the antenna directivities, $\beta_k$ is the distance pathloss exponent, $\xi_{n,k}$ is the complex path coefficient, $\phi_n$ is the \ac{AoD} defined from the antenna broadside, and $\tau^{\tt{A}}_n$ and $\tau^{\tt{D}}_n$ are the arrival and departure delays, respectively. 

Since the same stream of pilot symbols is received across antennas, we write the received signal at each sub-band as
\begin{equation}
    \pmb{Y}_k = \pmb{H}_k \diag(\pmb{s}_k) + \pmb{W}_k,
\end{equation}
where $\pmb{s}_k \in \mathbb{C}^{Q\times1}$ are the transmitted QAM-modulated symbols across the subcarriers, and the noise follows $\ve(\pmb{W}_k) \sim \mathcal{CN}(\pmb 0, \sigma^2_k \pmb I)$ , where $\sigma_k^2=\NF \cdot N_0\Delta f_k$, $N_0$ is the thermal noise power spectral density, and $\NF$ is the noise figure.

\subsection{Problem Formulation}

Note that $\pmb{Y}_k$ can be rewritten in vector form as 
\begin{equation}
    \ve(\pmb{Y}_k) = 
    \sum_{n=1}^N 
    g_{n,k}
    \left(\pmb{s}_k
    \odot
    \pmb{a}^{\tt{F}}_k(\tau_n)\right)
    \otimes
    \pmb{a}^{\tt{R}}_k(\theta_n)
    +\ve(\pmb{W}_k).
\end{equation}
We can think of the summation term due to the paths as a linear combination of the basis vectors
\begin{equation}
    \mathcal{B}_k = \left\{\pmb{s}_k \odot \pmb{a}^{\tt{F}}_k(\tau_n) \otimes \pmb{a}^{\tt{R}}_k(\theta_n)\right\}_{n=1}^N,
\end{equation}
each corresponding to a \ac{ToA}-\ac{AoA} pair. In order to resolve the paths via \ac{CS}, sparsity should be established in $\mathcal{B}_k$. We can do so by creating a ``fat" dictionary matrix $\pmb{A}_k=\pmb{A}^{\tt{F}}_k \otimes \pmb{A}^{\tt{R}}_k$ using very fine delay and angle grids to represent the basis vectors, which can be expressed by
\begin{equation}
\label{AF}
    \pmb{A}^{\tt{F}}_k = 
    \cat_2\left\{
    \pmb{s}_k
    \odot
    \pmb{a}^{\tt{F}}_k(\tau)
    \right\}_{\tau\in \mathcal{T}}
    \in \mathbb{C}^{Q\times L^{\tt{F}}},
\end{equation}
\begin{equation}
\label{AR}
    \pmb{A}^{\tt{R}}_k = 
    \cat_2\left\{
    \pmb{a}^{\tt{R}}_k(\theta)
    \right\}_{\theta\in \Theta}
    \in \mathbb{C}^{M\times L^{\tt{R}}},
\end{equation}
and the grids are given by
\begin{equation}
    \mathcal{T} = \left\{
        {i\tau_{\tt{max}}}/\left({L^{\tt F}-1}\right) \,\middle|\, i=0,1,\dots,L^{\tt F}-1
    \right\},
\end{equation}
\begin{equation}
    \Theta = \left\{
        -{\pi}/{2} + {i\pi}/\left({L^{\tt R}-1}\right) \,\middle|\, i=0,1,\dots,L^{\tt R}-1
    \right\}.
\end{equation}
where $L^{\tt{F}}$ and $L^{\tt{R}}$ are the grid sizes of $\mathcal{T}$ and $\Theta$, respectively.

Now, we can express $\pmb{Y}_k$ in the following form:
\begin{equation}
\label{sparsevecYk}
    \ve(\pmb{Y}_k) \approx \pmb{A}_k\pmb{x}_k + \ve(\pmb{W}_k),
\end{equation}
where $\pmb{x}_k$ is the \textit{sparse sensing vector} of the $k^\text{th}$ sub-band, and the approximation error is due to the discretization error from using grid search. We can think of $\pmb{x}_k$ as activating some sparse columns in the fine matrix $\pmb{A}_k$, each corresponding to a delay-angle pair closest to $\{\theta_n,\tau_n\}$. So, ideally, the value of each non-zero element of $\pmb{x}_k$ should approximate the corresponding $g_{n,k}$.
We can further exploit the Kronecker structure to speed up computations by rewriting \eqref{sparsevecYk} in the following form
\begin{equation}
    \pmb{Y}_k = \pmb{A}^{\tt{R}}_{k} \ve^{-1}(\pmb{x}_k) {\pmb{A}^{\tt{F}}_{k}}^T
    +\pmb{W}_k.
\end{equation}

To localize the targets, we notice the following:
As the \acp{ToA} and \acp{AoA} are expected to be similar across sub-bands due to the physical environment, each $\pmb{x}_k$ should activate the same dictionary columns and hence have the same indices with non-zero elements. However, since the signal power varies across sub-bands, the values of those non-zero elements may vary. 
Thus, we want to enforce \textit{structured sparsity}, which jointly optimizes for sparsity within each sub-band and preserves the same support locations across sub-bands. Ideally, we would need a constraint stating that each row of $\pmb{X}$ contains either all zeros or non-zero values,
but we relax the problem for tractability by using certain norms as the minimization objective instead of using such constraints. To independently obtain the sparsest solution for each sub-band, an $\ell_1$-norm objective is typically proposed \cite{donoho2006most}, so if we consider the multiband sparse sensing matrix $\pmb{X} = \cat_2\left\{\pmb{x}_k\right\}_{k=1}^K$, one optimization objective to promote structured sparsity in $\pmb{X}$ is to minimize the $\ell_{2,1}$-mixed norm of $\pmb{X}$, defined as  
\begin{equation}
    \norm{\pmb{X}}_{2,1} = \sum_{\ell=1}^{L^{\tt{F}}L^{\tt{R}}} \sqrt{\sum_{k=1}^{K}\bigl|\left[\pmb{X}\right]_{\ell,k}\bigr|^2 },
\end{equation}
which preserves the total energy of a path under any energy redistribution across sub-bands. We hence propose the following optimization problem for target localization:
\begin{equation}
\label{CMSproblem}
\begin{aligned}
(\mathcal{P}_{\tt{CMS}}):
\begin{cases}
\min\limits_{\pmb{X}} & 
\norm{
\pmb{X}
}_{2,1}
\\
\textrm{s.t.} 
& 
\norm{
{\pmb{Y}}_k - \pmb{A}^{\tt{R}}_{k} \ve^{-1}(\pmb{x}_k) {\pmb{A}^{\tt{F}}_{k}}^T
}_F
\leq \varepsilon_k, \forall k \\
\end{cases}
\end{aligned}
\end{equation}
where $\varepsilon_k$ are the denoise thresholds.

\section{Estimation Algorithm}
\label{sec:estimation-algorithm}
\subsection{Target Localization}
To solve the target localization problem posed in \eqref{CMSproblem}, we can use \ac{ADMM} to decompose the solving process into simpler updates. 
First, we normalize $\pmb{Y}_k$ by
$\tilde{\pmb{Y}}_k = \pmb{Y}_k/\norm{\pmb{Y}_k}_{F,2}$
to prevent signal power from impacting $\varepsilon_k$ due to scaled reconstruction errors $\ve(\pmb{H}_k \diag(\pmb{s}_k)) - \pmb{A}_k\pmb{x}_k$ for different signal powers even at the same \ac{SNR}.
Then, following the structure in \cite{boyd2011distributed}, we can introduce auxiliary variables $\pmb{Z}_{1\dots K}$ and rewrite the problem in the form
\begin{equation}
\label{CSproblemADMM}
\begin{aligned}
(\mathcal{P}_{\tt{ADMM}}):
\begin{cases}
\min\limits_{\pmb{X},\pmb{Z}_{1\dots K}}
& 
\norm{
\pmb{X}
}_{2,1}
\\
\textrm{s.t.} 
& \pmb{Z}_k \in \mathcal{C}_k,\\
& \pmb{Z}_k = \pmb{A}^{\tt{R}}_{k} \ve^{-1}(\pmb{x}_k) {\pmb{A}^{\tt{F}}_{k}}^T, \forall k \\
\end{cases}
\end{aligned}
\end{equation}
where $\mathcal{C}_k = \big\{ \pmb{Z}_k\,\big|\,\lVert \tilde{\pmb{Y}}_k - \pmb{Z}_k \rVert_F \leq \varepsilon_k\big\}$ is the convex feasible set of $\pmb{Z}_k$ for reconstruction fidelity. Hence, the augmented Lagrangian is given by
\begin{equation}
\begin{aligned}
\mathcal{L}_{\rho}\left(\pmb{X},\pmb{Z}_{1\dots K},\pmb{V}_{1\dots K}\right) 
&= 
\norm{
\pmb{X}
}_{2,1} + \sum_{k=1}^K \mathcal{I}_k(\pmb{Z}_{k}) \\
&+
\sum_{k=1}^K\trace (\pmb{V}^T_k(\pmb{A}^{\tt{R}}_{k} \ve^{-1}(\pmb{x}_{k} ){\pmb{A}^{\tt{F}}_{k}}^T - \mathbf{Z}_k)) \\
&+
\sum_{k=1}^K \frac{\rho_k}{2} 
\norm{
\pmb{A}^{\tt{R}}_{k} \ve^{-1}(\pmb{x}_{k} ){\pmb{A}^{\tt{F}}_{k}}^T - \mathbf{Z}_k
}_F^2,
\end{aligned}
\end{equation}
where $\rho_k$ is the per-sub-band \ac{ADMM} penalty factor, $\pmb{V}_{1\dots K}$ contains the Lagrangian dual variables, and each feasible set $\mathcal{C}_k$ can be enforced by the indicator function \cite{boyd2011distributed} \cite{parikh2014proximal} 
\begin{equation}
    \mathcal{I}_k(\pmb{Z}_k) = 
    \begin{cases}
    0 & \text{if } \pmb{Z}_k \in \mathcal{C}_k,\\
    \infty & \text{otherwise}.
    \end{cases}
\end{equation}

We then explain our variable update policies. First, we update $\pmb{X}$ while freezing $\pmb{Z}_{1\dots K}$ and $\pmb{V}_{1\dots K}$. From \cite{boyd2011distributed}, in this step, we want to find $\pmb{X}$ that minimizes the augmented Lagrangian. The part of the augmented Lagrangian that we care about is only the first, third and fourth terms. Since the $\ell_{2,1}$-mixed norm term is not smooth (i.e., having continuous gradients) while the third and fourth terms are, we can use the proximal gradient descent method \cite{parikh2014proximal}. We start by computing the gradient for the smooth terms:
\begin{equation}
\begin{aligned}
\nabla_{\pmb{x}_k} \mathcal{L}^{\tt{smth}}_{\rho}&= 
    {\pmb{A}^{\tt{R}}_k}^H \pmb{V}_k \overline{\pmb{A}^{\tt{F}}_k}\\
    &+
    \rho_k {\pmb{A}^{\tt{R}}_k}^H
    (\pmb{A}^{\tt{R}}_k\ve^{-1}{(\pmb{x}_k)}{\pmb{A}^{\tt{F}}_k}^T-\pmb{Z}_k)\overline{\pmb{A}^{\tt{F}}_k}.
\end{aligned}
\end{equation}
To speed up computations, we can scale the dual variables. Let $\pmb{U}_k = \pmb{V}_k/\rho_k$, we can rewrite the gradient as
\begin{equation}
\label{gradientSmoothSimple}
    \nabla_{\pmb{x}_k} \mathcal{L}^{\tt{smth}}_{\rho} = 
    \rho_k {\pmb{A}^{\tt{R}}_k}^H(\pmb{A}^{\tt{R}}_k\ve^{-1}{(\pmb{x}_k)}{\pmb{A}^{\tt{F}}_k}^T-\pmb{Z}_k+\pmb{U}_k)\overline{\pmb{A}^{\tt{F}}_k}.
\end{equation}
Denoting $\nabla_{\pmb{X}}\mathcal{L}^{\tt{smth}}_\rho = \cat_2\left\{\nabla_{\pmb{x}_k} \mathcal{L}^{\tt{smth}}_{\rho} \right\}_{k=1}^K$, we update the target variable by applying a proximal operator of the objective function onto the smooth gradient descent result:
\begin{equation}
\label{Xupdate}
    \pmb{X}^{(i+1)} = \prox_{\gamma \norm{ \cdot}_{2,1}
    }\left( \pmb{X}^{(i)} - \gamma\nabla_{\pmb{X}}\mathcal{L}^{\tt{smth}}_\rho
    \right).
\end{equation}
In our case, we use group soft thresholding, which is a common proximal operator for the $\ell_{2,1}$-mixed norm: 
\begin{equation}
\label{proxX}
    \prox_{\gamma \norm{\cdot}_{2,1}}(\pmb{C}) = 
    \left[
    \max\left(0,1-{\gamma}/
    {\bigl \lVert {\left[\pmb{C}\right]_{\ell,:}} \bigr \rVert_2}
    \right)\left[\pmb{C}\right]_{\ell,:}
    \right]_{\ell=1}^{N_C},
\end{equation}
where $\pmb{C}$ is some matrix with $N_C$ rows.
Since the $\nabla_{\pmb{x}_k} \mathcal{L}_{\rho}'$ has the form $\rho_k\left(\pmb{A}^H_k\pmb{A}_k\right) \pmb{x}_k+ \pmb{b}$, the step size is conservatively determined from the Lipschitz's constraint:
\begin{equation}
\label{gamma}
    \gamma = \left(\max_k \left( \rho_k\norm{ \pmb{A}^{\tt{F}}_k }_2^2
    \norm{ \pmb{A}^{\tt{R}}_k }_2^2 \right)\right)^{-1}.
\end{equation}

Next we update $\pmb{Z}_{1\dots K}$. From \cite{boyd2011distributed}, in this step, we apply the proximal operator as follows: 
\begin{equation}
\label{Zupdate}
    \pmb{Z}_{k}^{(i+1)} = 
    \prox_{\mathcal{I}_k(\cdot)} \left(
    \pmb{A}^{\tt{R}}_k\ve^{-1}{\left(\pmb{x}_k^{(i+1)}\right)}{\pmb{A}^{\tt{F}}_k}^T + 
    \pmb{U}_{k}^{(i)}\right) , \forall k.
\end{equation}
From \cite{parikh2014proximal}, since we converted the inequality constraints using an indicator function, the proximal operator reduces to a Frobenius norm ball projection centered at $\tilde{\pmb{Y}}_k$ with radius of $\varepsilon_k$:
\begin{equation}
\label{proxZ}
    \prox_{\mathcal{I}_k(\cdot)}(\pmb{C}) =
    \begin{cases}
        \pmb{C} &\text{if } 
        \pmb{C} \in \mathcal{C}_k,\\
        \tilde{\pmb{Y}}_k + \varepsilon_k\cdot\frac{\pmb{C}-\tilde{\pmb{Y}}_k}{\lVert\pmb{C}-\tilde{\pmb{Y}}_k\rVert_F} &\text{otherwise}.
    \end{cases}
\end{equation}
After that, we update $\pmb{U}_{1\dots K}$ as described in \cite{boyd2011distributed}:
\begin{equation}
\label{Uupdate}
    \pmb{U}_{k}^{(i+1)} = 
    \pmb{U}_{k}^{(i)} + 
    \pmb{A}^{\tt{R}}_{k} \ve^{-1}\left(\pmb{x}^{(i+1)}_{k} \right){\pmb{A}^{\tt{F}}_{k}}^T - \pmb{Z}^{(i+1)}_k, \forall k.
\end{equation}

Finally, we combine the contributions from all sub-bands through arithmetic averaging. The output profile is given by 
\begin{equation}
\label{Xprofile}
X(\theta,\tau) = \left[
    \frac{1}{K}
    \sum_{k=1}^K
    \ve^{-1}(|\pmb{x}_k^{\tt{f}}|)
    \right]_{ij},  \forall \{\theta_i,\tau_j\}\in\Theta\times \mathcal{T}.
\end{equation}
where $\pmb{x}^{\tt{f}}_k$ is the final $\pmb{x}_k$ output. The support locations and the number of them are jointly estimated through performing peakfinding with a threshold on the normalized output profile:
\begin{equation}
\label{peakfinding}
\{\hat{\theta}_n,\hat{\tau}_n\} \in
    \argmax_{\theta,\tau}
    \left(
    \max\left(r_{\tt{pf}},
    \frac{X(\theta,\tau)}
    {\max X(\theta,\tau)}\right)\right),
\end{equation}
where $r_{\tt{pf}}$ is the pre-chosen peakfinding threshold. We denote the number of estimated peaks by $\hat{N}$.

\subsection{Adaptive Learning Parameters}

To maintain diversity between sub-bands and make the algorithm more efficient, we consider three aspects of adaptive learning parameters: adaptive $\varepsilon_k$, adaptive $\rho_k$, and automatic halt of the algorithm. We initialize $\varepsilon_k$ by $\varepsilon_k^{\tt{i}} = \alpha \lVert \tilde{\pmb{Y}}_k \rVert_F$. Then, by observing the constraint residuals, we can loosen or tighten $\varepsilon_k$ in the iterations through
\begin{equation}
\label{epsilonupdate}
	\varepsilon_k^{(i+1)} = \varepsilon_k^{(i)} + \mu_\varepsilon \left(
	\norm{\tilde{\pmb{Y}}_k - \pmb{A}^{\tt{R}}_{k} \ve^{-1}\left(\pmb{x}_k^{(i+1)}\right) {\pmb{A}^{\tt{F}}_{k}}^T}_F - \varepsilon_k^{(i)} \right).
\end{equation}
where $\rho_k$ balances the primal and dual updates. Following the scheme from \cite{boyd2011distributed}, in each iteration, we first calculate the per-sub-band primal residual 
\begin{equation}
\label{pres}
{\textrm{p-res}}_k^{(i+1)} = \norm{
\pmb{A}^{\tt{R}}_k \ve^{-1}\left(\pmb{x}^{(i+1)}_k\right){\pmb{A}^{\tt{F}}_k}^T
-\pmb{Z}^{(i+1)}_k}_F
\end{equation}
and the dual residual 
\begin{equation}
\label{dres}
    {\textrm{d-res}}_k^{(i+1)} = 
\rho_k^{(i)}\norm{\pmb{Z}_k^{(i+1)}-\pmb{Z}_k^{(i)}}_F.
\end{equation}
We then update $\rho_k$ in each iteration by
\begin{equation}
\label{rhoupdate}
    \rho_k^{(i+1)} = 
    \begin{cases}
    \rho_k^{(i)} \mu_\rho & \text{if } {\textrm{p-res}}_k^{(i+1)}>r_\rho{\textrm{d-res}}_k^{(i+1)},\\
        \rho_k^{(i)} \mu^{-1}_\rho & \text{if } {\textrm{d-res}}_k^{(i+1)}>r_\rho{\textrm{p-res}}_k^{(i+1)},\\
    \end{cases}
\end{equation}
where $r_\rho$ is pre-chosen. We finally rescale $\pmb{U}_k$ and recalculate $\gamma$ after this step. 

To decide when to halt \ac{ADMM} automatically, we consider solution feasibility and stability by computing the primal and dual residuals \cite{boyd2011distributed}. In short, convergence is declared when
\begin{equation}
\label{convergencecrit}
\begin{aligned}
    \textrm{p-res} &= \norm{[\textrm{p-res}_k]_{k=1}^K}_2 \leq \epsilon_{\tt prim}, \\
    \textrm{d-res} &= \norm{[\textrm{d-res}_k]_{k=1}^K}_2 \leq \epsilon_{\tt dual},
\end{aligned}
\end{equation}
where $\epsilon_{\tt{prim}}$ and $\epsilon_{\tt{dual}}$ are the primal and dual tolerances, respectively. Similar to the residuals, we first calculate the per-sub-band tolerances as outlined in \cite{boyd2011distributed} using specified absolute and relative tolerances $\epsilon_{\tt{abs}}$ and $\epsilon_{\tt{rel}}$, respectively, then aggregate across sub-bands using the $\ell_2$-norm approach.
Algorithm \ref{alg:admm} provides a summary of the ADMM-CMS algorithm.
\begin{algorithm}[!t]
\caption{ADMM-CMS}
\label{alg:admm}
\KwIn{ $ \{\pmb{Y}_{k} ,\pmb{s}_{k} \}_{k=1}^{K} $ }
\KwOut{ $ \{\hat\theta_n, \hat\tau_n\}_{n=1}^{\hat N}$, $\hat{N}$}
$ \tilde{\pmb{Y}}_k \gets \pmb{Y}_k/\norm{\pmb{Y}_k}_{F,2}, \forall k $ \\
Compute $ \pmb{A}^{\tt{R}}_k, \forall k$ by \eqref{AR} \\
Compute $ \pmb{A}^{\tt{F}}_k, \forall k$ by \eqref{AF} \\
Initialize $\{ \pmb{X}, \pmb Z_k, \pmb U_k\} \gets 0 $, $ \varepsilon_k \gets \varepsilon^{\tt{i}}_k $, $ \rho_k \gets \rho^{\tt{i}}, \forall k $ \\

Initialize $ \gamma $ by \eqref{gamma} \\
    \For{$ i = 1 $ \textbf{to} $ {\tt{max\_iter}} $}{
    Compute $ \nabla_{\pmb{x}_k} \mathcal{L}^{\tt{smth}}_\rho,\forall k $ by \eqref{gradientSmoothSimple} \\
        
        $\nabla_{\pmb{X}}\mathcal{L}^{\tt{smth}}_\rho \gets \cat_2\left\{\nabla_{\pmb{x}_k} \mathcal{L}^{\tt{smth}}_\rho\right\}_{k=1}^K$ \\
        Update $ \pmb{X} $ by \eqref{Xupdate} \eqref{proxX} \\
        \For{$ k = 1 $ \textbf{to} $ K $}{
            Update $ \pmb{Z}_k $ by \eqref{Zupdate} \eqref{proxZ} \\
            Update $ \pmb{U}_k $ by \eqref{Uupdate} \\
            Update $ {\textrm{p-res}}_k $ by \eqref{pres}\\
            Update $ {\textrm{d-res}}_k $ by \eqref{dres}\\
        	Update $ \rho_k $ by \eqref{rhoupdate}\\
        	$\pmb{U}_k \gets \pmb{U}_k/\mu_\rho \text{ or } \pmb{U}_k/\mu^{-1}_\rho $ depending on \eqref{rhoupdate} \\      
        	Update $ \varepsilon_k $ by \eqref{epsilonupdate}\\  
        }
        Compute $\gamma$ by \eqref{gamma} \\
        Compute $ \epsilon_{\tt{prim}} $ and $ \epsilon_{\tt{dual}} $ as outlined in \cite{boyd2011distributed}\\
        \If{\eqref{convergencecrit}}{
            \textbf{break}
        }
    }
    Estimate $ \{\hat\theta_n, \hat\tau_n\},\forall n$ and $\hat{N}$ by \eqref{Xprofile} \eqref{peakfinding} \\
\end{algorithm}

\section{Simulation Results}
\label{sec:simulation-results}
\subsection{Comparison Benchmark}

A Bartlett-type BF-based approach is used in \cite{raviv2024multi}. To adapt this benchmark to multiband sensing, we calculate the output power profile for all $\{\theta_i,\tau_j\}\in\Theta\times \mathcal{T}$ by
\begin{equation}
    P(\theta,\tau) =  
    \frac{1}{K}\sum_{k=1}^K
    \left|
    \left(\pmb{s}_k \odot
    \pmb{a}^{\tt{F}}_k(\tau_j)\right)^T
    \pmb{Y}_k^H 
    \left(\pmb{a}^{\tt{R}}_k(\theta_i)\right)
    \right|,
\end{equation}
and then perform a peakfinding similar to \eqref{peakfinding} for $P(\theta,\tau)$.

\subsection{Metrics}
For the \ac{RMSE} metric, we only consider errors of the true and estimated point pairs that are optimally matched without replacement using the Hungarian algorithm. This is because we are doing joint estimation and detection, so the number of estimated targets may be different from the actual number of targets. We evaluate the \ac{RMSE} for both \acp{AoA} and \acp{ToA}.

Another metric that we consider is the \ac{SRP}, which focuses only on localization performance. It is defined as the fraction of all Monte Carlo simulations that has $\hat{N} = N$, $|\hat\theta_n-\theta_n| \leq 1^\circ, \forall n$, and $|\hat\tau_n-\tau_n| \leq 1 \text{ ns}, \forall n$ satisfied simultaneously.

\subsection{Simulation Implementation}
For simplicity, we assume omnidirectional antennas, hence $u_k^{\tt{T}}(\phi)=u_k^{\tt{R}}(\theta) = 1$. Table \ref{table:parameters} provides all simulation parameters, where $\tt{num\_MC}$ is the number of Monte Carlo iterations. Note that in the proposed numerology of FR3, we expect higher bandwidths at higher sub-bands \cite{bazzi2025uppermidbandspectrum6g}. We predefine the \acp{AoA}, \acp{AoD}, and \acp{ToA} of the \ac{Tx}, \ac{Rx}, and scatterers. These sensing parameters are intentionally generated off-grid for better visualization of the \ac{RMSE}. The test variable is $P^{\tt{T}}$.

\begin{table}[!t]
\renewcommand{\arraystretch}{1.3}
\caption{Simulation Parameters}
\label{table:parameters}
\centering
\newcolumntype{Y}{>{\centering\arraybackslash}m{64.25pt}}
\begin{tabularx}{\columnwidth}{|c|Y||c|Y|}
\hline
\rowcolor{gray10pct} \multicolumn{4}{|c|}{\bfseries System Parameters} \\
\hline
$N$ & 2 & $M$ & 60 \\
\hline
$Q$ & 100 & $K$ & 2 \\
\hline
$f_k$ & $\{7,10\}$ GHz (FR3) & $\Delta f_k$ & $\{1800,3000\}$ kHz \\
\hline
$\beta_k$ & 1.34, $\forall k$ \cite{shakya2024meas} & Modulation & 4-QAM \\
\hline
$d_k$ & $\lambda_k/2$ & $P^{\tt{T}}$ & -70 to -20 dBm \\
\hline
$N_0$ & -174 dBm/Hz & $\NF$ & 7 dB \\
\hline
$\tau_{\tt{max}}$ & $200$ ns & $\xi_{n,k}$ & $\{1,5\},\forall k$ (Fig. \ref{fig:srp-pt}); $1,\forall n,k$ (Fig. \ref{fig:rmse-pt})\\
\hline
\rowcolor{gray10pct} \multicolumn{4}{|c|}{\bfseries Sensing Parameters} \\
\hline
$\tau_n$ & $\{40.03,45.08\}$ ns & $\theta_n$ & $\{0,45.2\}^\circ$ \\
\hline
\rowcolor{gray10pct} \multicolumn{4}{|c|}{\bfseries Algorithm Parameters} \\
\hline
$\tt{max\_iter}$ & $10^4$ & $\tt{num\_MC}$ & 900 \\
\hline
$L^{\tt{F}}$ & 201 ($1$ ns steps) & $L^{\tt{R}}$ & 181 ($1^\circ$ steps) \\
\hline
$\alpha$ & 0.01 & $\mu_\varepsilon$ & 0.3 \\
\hline
$\rho^{\tt{i}}$ & 0.2 & $\mu_\rho$ & 1.001 \\
\hline
$r_\rho$ & 10 & $r_{\tt{pf}}$ & 0.2 (Fig. \ref{fig:srp-pt}); 0.02 (Fig. \ref{fig:rmse-pt})  \\
\hline
$\epsilon_{\tt{abs}}$ & $10^{-8}$ & $\epsilon_{\tt{rel}}$ & $10^{-5}$ \\
\hline
\end{tabularx}
\vspace{-10pt}
\end{table}

\subsection{Results and Discussion}

\begin{figure}[!t]
\centering
\includegraphics[width=3.5in]{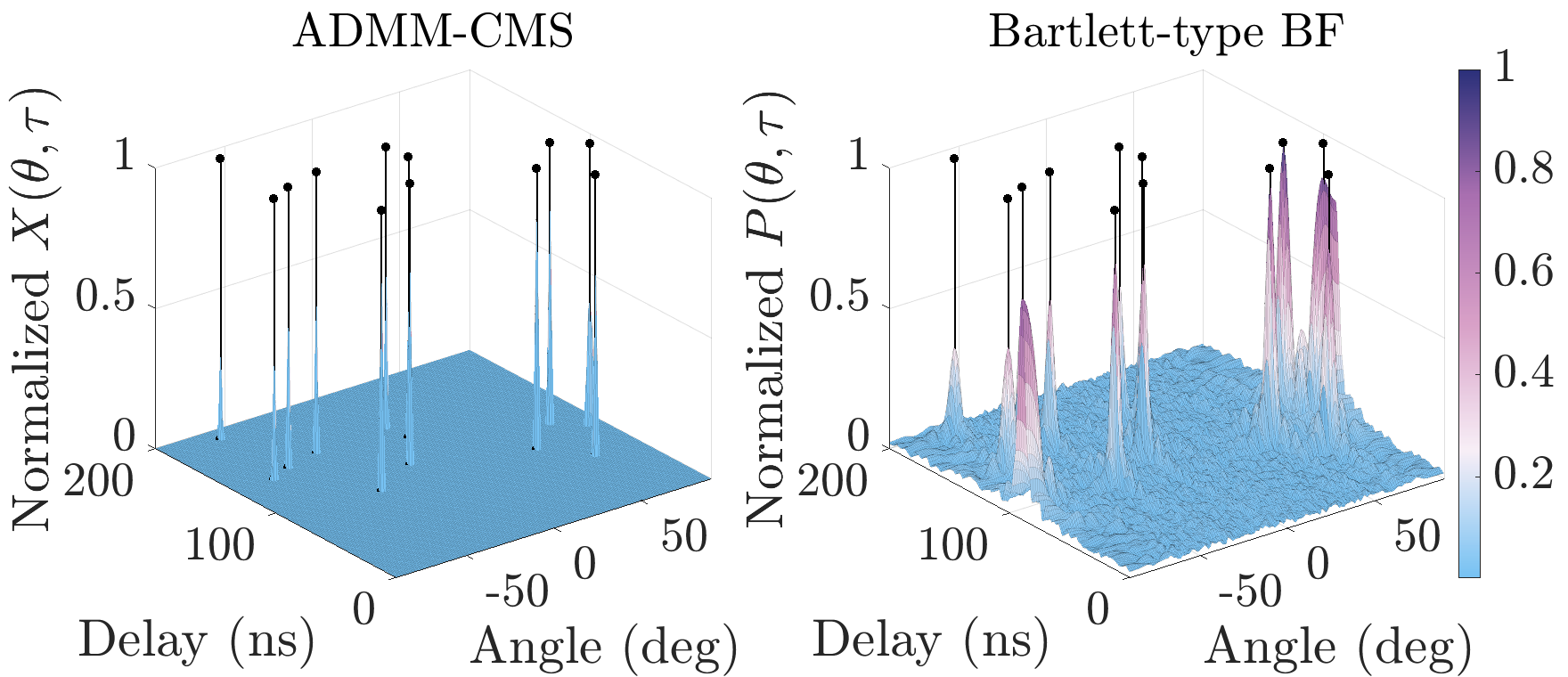}
\caption{Sample normalized $X(\theta,\tau)$ of our ADMM-CMS and $P(\theta,\tau)$ of the Bartlett-type \ac{BF} benchmark. The black lines correspond to the true \acp{AoA} and \acp{ToA}, and $P^{\tt{T}} = -25 \textrm{ dBm}$.}
\label{fig:profile}
\vspace{-10pt}
\end{figure}

From Fig. \ref{fig:profile}, our ADMM-CMS produces much sharper and less noisy peaks compared to the Bartlett-type \ac{BF} benchmark under the same system parameters, which implies improved detection and resolution. 

\begin{figure}[!t]
\centering
\includegraphics[width=3.5in]{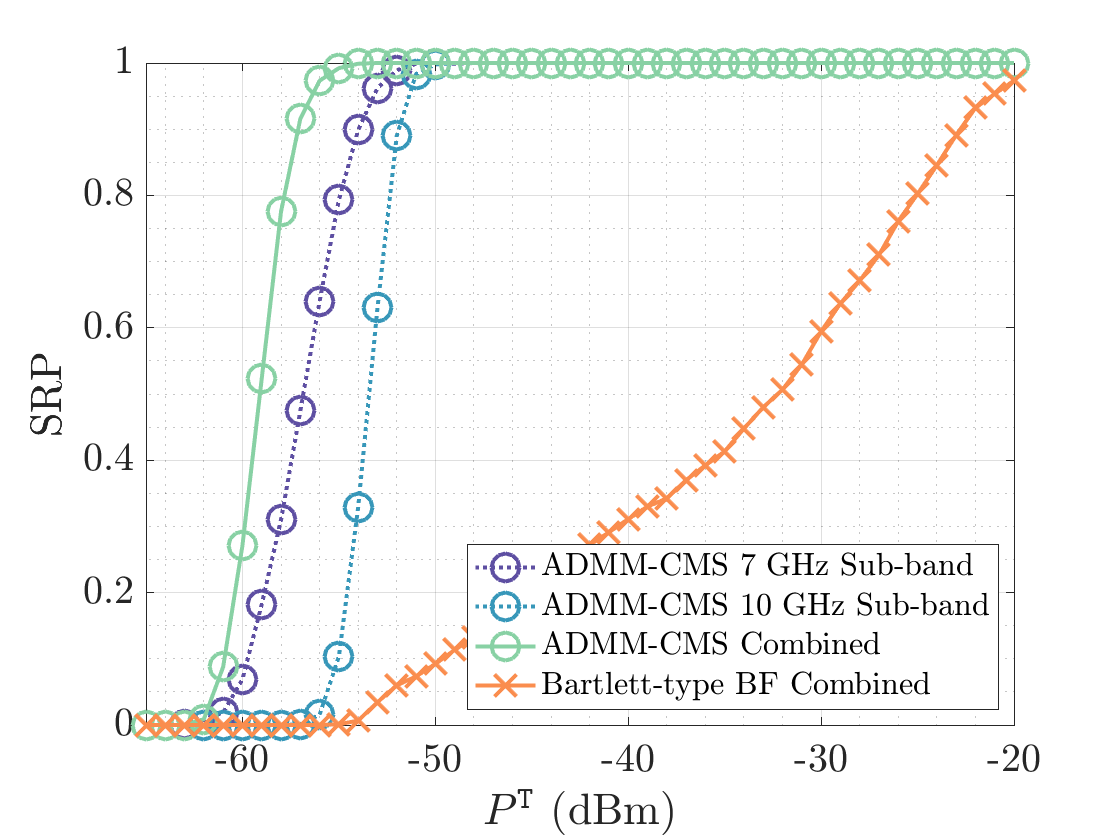}
\vspace{-20pt}
\caption{\ac{SRP} vs. $P^{\tt{T}}$ for one \ac{LoS} path and one scatterer for both ADMM-CMS and the Bartlett-type \ac{BF} benchmark.}
\label{fig:srp-pt}
\vspace{-10pt}
\end{figure}

As a result, from Fig. \ref{fig:srp-pt}, our ADMM-CMS achieves significantly better \ac{SRP} performance especially in the high \ac{SNR} region (at $P^{\tt{T}} = -62$ dBm and above) as compared to the Bartlett-type \ac{BF} benchmark. For example, for an \ac{SRP} of 0.9, the ADMM-CMS requires only about -57 dBm as opposed to -23 dBm for the Bartlett-type \ac{BF} benchmark. One contributor to this large gain is that even at high \ac{SNR}, the $P(\theta,\tau)$ profile for the Bartlett-type \ac{BF} benchmark contains sinc-shaped artifacts along the axes corresponding to the estimated \acp{ToA} and \acp{AoA}. Moreover, we can also see significant \ac{SRP} gains of ADMM-CMS compared to performing \ac{CS} on the constituent sub-bands. For instance, at the same \ac{SNR} of $P^{\tt{T}} = -56$ dBm, the combined case has an \ac{SRP} of 0.97, while those for lower and higher sub-bands are only 0.65 and 0.02, respectively. 

\begin{figure}[!t]
\centering
\includegraphics[width=3.5in]{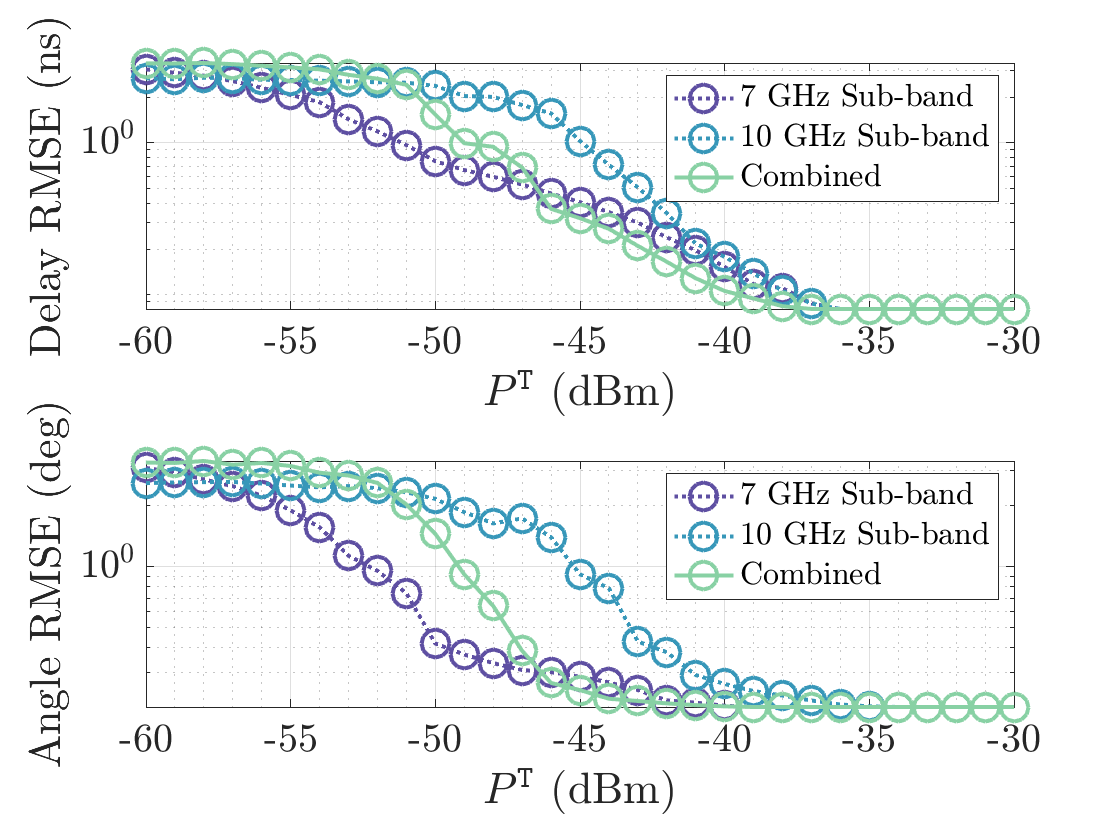}
\vspace{-20pt}
\caption{Delay and angle \ac{RMSE} vs. $P^{\tt{T}}$ for one \ac{LoS} path and one scatterer for ADMM-CMS.}
\label{fig:rmse-pt}
\vspace{-10pt}
\end{figure} 

From Fig. \ref{fig:rmse-pt}, we can see that the ADMM-CMS also achieves a better \ac{RMSE} than performing \ac{CS} on the constituent sub-bands in the high \ac{SNR} region (at $P^{\tt{T}} = -46$ dBm and above). For instance, at -41 dBm, the combined case has a delay \ac{RMSE} of about 0.13 ns while those for lower and higher sub-bands are around 0.20 and 0.21 ns, respectively.
However, in Fig. \ref{fig:rmse-pt}, we can see that the start of the high \ac{SNR} region (signaled by the end of the waterfall region) for the combined case lies between that of the constituent sub-bands, while in Fig. \ref{fig:srp-pt}, the combined case shows an earlier start of the transition region compared to the sub-bands.

\section{Conclusion and Future Work}
\label{sec:conclusion-and-future-work}
This paper has shown that with the same 100 received pilot symbols across 60 antennas, the proposed ADMM-CMS method achieves substantially better performance than the benchmark approaches. First, ADMM-CMS produces sharper delay–angle resolution and stronger denoising than Bartlett \ac{BF}, leading to higher \ac{SRP}. For example, to attain an \ac{SRP} of 0.9, ADMM-CMS achieves a 34 dB gain in $P^{\tt{T}}$ as compared to Barlett \ac{BF}, which demonstrates that enforcing structured sparsity across sub-bands effectively suppresses noise and false artifacts, improving target detection. 
Second, ADMM-CMS outperforms compressed sensing applied independently to constituent sub-bands using the same \ac{ADMM} framework in terms of estimation accuracy and target detection. For instance, under $P^{\tt{T}} = -41 \textrm{ dBm}$, ADMM-CMS reduces the delay \ac{RMSE} by 34.46\% and 40.76\%, respectively, compared to separately employing the constituent 7 and 10 GHz sub-bands, which shows that leveraging multiple sub-bands greatly enhances localization accuracy.

Together, these findings establish ADMM-CMS as a practical enabler of \ac{ISAC} in FR3, offering both target detectability and spatial resolution advantages critical for future 6G applications such as the Internet of Things, smart cities, and autonomous driving.
For future work, a more realistic channel model with sub-band frequency-dependent dense multipath components  \cite{molisch2022dmc, ying2026site} and identical antenna spacing across all sub-bands can be explored. 

\section*{Acknowledgments}
\label{sec:acknowledgments}
This work is supported by Tamkeen under the Research Institute NYUAD grant CG017, and the New York University (NYU) WIRELESS Industrial Affiliates Program. 

\bibliographystyle{IEEEtran}
\bibliography{refs}

@ARTICLE{chafii2023twelve,
  author={Marwa Chafii and Lina Bariah and Sami Muhaidat and Merouane Debbah},
  journal={IEEE Communications Surveys \& Tutorials}, 
  title={{Twelve Scientific Challenges for 6G: Rethinking the Foundations of Communications Theory}}, 
  year={2023},	
  volume={},
  number={},
  pages={1-1},
  doi={10.1109/COMST.2023.3243918}}

@INPROCEEDINGS{raviv2024multi,
  author={Raviv, Tomer and Kang, Seongjoon and Mezzavilla, Marco and Rangan, Sundeep and Shlezinger, Nir},
  booktitle={2024 IEEE 25th International Workshop on Signal Processing Advances in Wireless Communications (SPAWC)}, 
  title={{Multi-Frequency Upper Mid-Band Localization}}, 
  year={2024},
  volume={},
  number={},
  pages={736-740},
  doi={10.1109/SPAWC60668.2024.10694061}}

@INPROCEEDINGS{wu2011csnoisered,
  author={Wu, Dalei and Zhu, Wei-Ping and Swamy, M.N.S.},
  booktitle={{2011 IEEE 54th International Midwest Symposium on Circuits and Systems (MWSCAS)}}, 
  title={{A Compressive Sensing Method for Noise Reduction of Speech and Audio Signals}}, 
  year={2011},
  volume={},
  number={},
  pages={1-4},
  doi={10.1109/MWSCAS.2011.6026662}}

@article{boyd2011distributed,
  title={{Distributed optimization and statistical learning via the alternating direction method of multipliers}},
  author={Boyd, Stephen and Parikh, Neal and Chu, Eric and Peleato, Borja and Eckstein, Jonathan and others},
  journal={Foundations and Trends{\textregistered} in Machine learning},
  volume={3},
  number={1},
  pages={1--122},
  year={2011},
  publisher={Now Publishers, Inc.}
}

@article{donoho2006most,
  title={{For most large underdetermined systems of linear equations the minimal $\ell_1$-norm solution is also the sparsest solution}},
  author={Donoho, David L},
  journal={Communications on Pure and Applied Mathematics: A Journal Issued by the Courant Institute of Mathematical Sciences},
  volume={59},
  number={6},
  pages={797--829},
  year={2006},
  publisher={Wiley Online Library}
}

@article{parikh2014proximal,
  title={{Proximal algorithms}},
  author={Parikh, Neal and Boyd, Stephen and others},
  journal={Foundations and trends{\textregistered} in Optimization},
  volume={1},
  number={3},
  pages={127--239},
  year={2014},
  publisher={Now Publishers, Inc.}
}

@ARTICLE{molisch2022dmc,
  author={Jiang, Suying and Wang, Wei and Miao, Yang and Fan, Wei and Molisch, Andreas F.},
  journal={IEEE Open Journal of Antennas and Propagation}, 
  title={{A Survey of Dense Multipath and Its Impact on Wireless Systems}}, 
  year={2022},
  volume={3},
  number={},
  pages={435-460},
  doi={10.1109/OJAP.2022.3168400}}

@article{bazzi2025uppermidbandspectrum6g,
  author  = {Bazzi, Ahmad and Bomfin, Roberto and Mezzavilla, Marco and Rangan, Sundeep and Rappaport, Theodore S. and Chafii, Marwa},
  title   = {{Upper Mid-Band Spectrum for 6G: Vision, Opportunity and Challenges}},
  journal = {IEEE Communications Magazine},
  year    = {2026},
  volume  = {64},
  number  = {1},
  pages   = {206--212},
  month   = jan,
  doi     = {10.1109/MCOM.001.2500101}
}

@ARTICLE{shakya2024meas,
  author={Shakya, Dipankar and Ying, Mingjun and Rappaport, Theodore S. and Poddar, Hitesh and Ma, Peijie and Wang, Yanbo and Al-Wazani, Idris},
  journal={IEEE Open Journal of the Communications Society}, 
  title={{Comprehensive FR1(C) and FR3 Lower and Upper Mid-Band Propagation and Material Penetration Loss Measurements and Channel Models in Indoor Environment for 5G and 6G}}, 
  year={2024},
  volume={5},
  number={},
  pages={5192-5218},
  doi={10.1109/OJCOMS.2024.3431686}}

@article{chi2015atomicnorm,
  title={{Compressive two-dimensional harmonic retrieval via atomic norm minimization}},
  author={Chi, Yuejie and Chen, Yuxin},
  journal={IEEE Transactions on Signal Processing},
  volume={63},
  number={4},
  pages={1030--1042},
  year={2015},
  doi={10.1109/TSP.2014.2386293}
}

@article{liu2018crb,
  title={{Cram{\'e}r--Rao bounds for joint delay and angle estimation in compressive sensing radar}},
  author={Liu, An and Liu, Xiaoli and Zeng, Yong and Lau, Vincent K. N.},
  journal={IEEE Transactions on Signal Processing},
  volume={66},
  number={18},
  pages={4772--4787},
  year={2018},
  doi={10.1109/TSP.2018.2851397}
}

@article{Gong2023SAGE,
  title     = {{EM and {SAGE} Algorithms for {DOA} Estimation in the Presence of Unknown Uniform Noise}},
  author    = {Ming-Yan Gong and Bin Lyu},
  journal   = {Sensors},
  volume    = {23},
  number    = {10},
  pages     = {4811},
  year      = {2023},
  publisher = {MDPI},
  doi       = {10.3390/s23104811}
}

@ARTICLE{rap2019100,
  author={Rappaport, Theodore S. and Xing, Yunchou and Kanhere, Ojas and Ju, Shihao and Madanayake, Arjuna and Mandal, Soumyajit and Alkhateeb, Ahmed and Trichopoulos, Georgios C.},
  journal={IEEE Access}, 
  title={{Wireless Communications and Applications Above 100 GHz: Opportunities and Challenges for 6G and Beyond}}, 
  year={2019},
  volume={7},
  number={},
  pages={78729-78757},
  doi={10.1109/ACCESS.2019.2921522}}

@book{liberti1999smart,
  title={{Smart antennas for wireless communications: IS-95 and third generation CDMA applications}},
  author={Liberti, Joseph C and Rappaport, Theodore S},
  year={1999},
  chapter={8--10},
  publisher={Prentice Hall PTR}
}

@INPROCEEDINGS{Shakya2025icc,
    author={{D.Shakya \textit{et al.}}},
    booktitle={IEEE ICC 2025}, 
    title={{Urban Outdoor Propagation Measurements and Channel Models at 6.75  {GHz}  FR1(C) and 16.95  {GHz}  FR3 Upper Mid-Band Spectrum for 5G and 6G}}, 
    year={2025},
    volume={},
    number={},
    pages={1-6}}

@INPROCEEDINGS{Shakya2024gc2,
    author={{D. Shakya \textit{et al.}}},
    booktitle={IEEE GLOBECOM 2024}, 
    title={{Wideband Penetration Loss through Building Materials and Partitions at 6.75 GHz in FR1(C) and 16.95 GHz in the FR3 Upper Mid-band spectrum}}, 
    year={2024},
    volume={},
    number={},
    pages={1-6}}

@INPROCEEDINGS{Ying2025icc,
 	author={{M. Ying, D. Shakya, T. S. Rappaport, P. Ma, Y. Wang, I. Al-Wazani, Y. Wu, and H. Poddar}},
 	booktitle={IEEE ICC 2025},
 	title={{Upper Mid-Band Channel Measurements and Characterization at 6.75 GHz FR1(C) and 16.95 GHz FR3 in an Indoor Factory Scenario}},
 	year={2025},
 	volume={},
 	number={},
 	pages={1-6}}

@INPROCEEDINGS{Ted2025icc,
	author={{T. S. Rappaport, D. Shakya, and M. Ying}},
	booktitle={IEEE International Communications Conference (ICC) 2025},
	title={{Point data for site-specific mid-band radio propagation channel statistics in the indoor hotspot (InH) environment for 3GPP and Next Generation Alliance (NGA) channel modeling}},
	year={2025},
	volume={},
	number={},
	pages={1-6}}

@article{ying2026site,
  author  = {Ying, Mingjun and Shakya, Dipankar and Ma, Peijie and Qian, Guanyue and Rappaport, Theodore S.},
  title   = {Site-specific location calibration and validation of ray-tracing simulator {NYURay} at upper mid-band frequencies},
  journal = {npj Wireless Technology},
  year    = {2026},
  month   = mar,
  pages   = {1--18},
  doi     = {10.1038/s44459-025-00014-x}
}

\vspace{12pt}

\end{document}